\documentclass{aastex61}

\newcommand\aastex{AAS\TeX}

\received{July 1, 2016}
\revised{September 27, 2016}
\accepted{\today}

\submitjournal{ApJ}

\shorttitle{\aastex\ Galactic dynamics}
\shortauthors{van Putten}

\begin{document}

\title{Evidence for galaxy dynamics tracing background cosmology below the de Sitter scale of acceleration}

\correspondingauthor{Maurice H.P.M. van Putten}
\email{mvp@sejong.ac.kr}

\author[0000-0002-0000-0000]{Maurice H.P.M. van Putten}
\affil{Physics and Astronomy, Sejong University\\
209 Neungdong-ro, Gwangjin-Gu \\
Seoul 143-747, Korea}
\begin{abstract}
Galaxy dynamics probes weak gravity at accelerations below the de Sitter scale of acceleration $a_{dS}=cH$, where $c$ is the velocity of light and $H$ is the Hubble parameter. Low and high redshift galaxies hereby offer a novel probe of weak gravity in an evolving cosmology, satisfying $H(z)=H_0\sqrt{1+\omega_m(6z+12z^2+12z^3+6z^4+(6/5)z^5)}/(1+z)$ with baryonic matter content $\omega_m$ sans tension to $H_0$ in surveys of the Local Universe.
Galaxy rotation curves show anomalous galaxy dynamics in weak gravity $a_N<a_{dS}$ across a transition 
radius $r_t = 4.7\,\mbox{kpc}\,M_{11}^{1/2}(H_0/H)^\frac{1}{2}$ in galaxies of mass $M=10^{11}M_\odot M_{11}$, where $a_N$ is the Newtonian acceleration based on baryonic matter content. 
We identify this behavior with a holographic origin of inertia from entanglement entropy, that introduces a $C^0$ onset
across $a_N=a_{dS}$ with asymptotic behavior described by a Milgrom parameter satisfying $a_0=\omega_0/2\pi$,
where $\omega_0=\sqrt{1-q}H$ is a fundamental eigenfrequency of the cosmological horizon.
Extending an earlier confrontation with data covering $0.003\lesssim a_N/a_{dS}\lesssim1$ at redshift $z\sim0$ in Lellie et al. (2016),
the modest anomalous behavior in the Genzel et al. sample at redshifts $0.854\le z\le 2.282$ is found to be mostly due to clustering $0.36\lesssim a_N/a_{dS}\lesssim1$ close to the $C^0$ onset to weak gravity and an increase of up to 65\% in $a_0$.
\end{abstract}


\section{Introduction}

Galaxy dynamics offers a unique view on weak gravitational interactions complementary to cosmology, showing weak gravitational repulsion in
surveys of the Local Universe \citep{rie98,per99} and probes of the Cosmic Microwave Background (CMB) \citep{ade13}. 
At accelerations below the de Sitter scale of acceleration 
\begin{eqnarray}
a_{dS}=cH, 
\label{EQN_adS}
\end{eqnarray}
where $c$ is the velocity of light and $H$ is the Hubble parameter, anomalous behavior is observed in accelerations that exceed Newtonian accelerations $a_N$ based on baryonic matter content inferred from the observed light \citep{fam12}. Presently, this is attributed to the presence of additional dark matter, enhancing gravitational attraction, or a modification of Newtonian dynamics, that may include relaxing the equivalence of inertial and gravitational mass associated with accelerations below (\ref{EQN_adS}) \citep{mil99,smo17}, where inertia varies with acceleration. 

The mass of the putative dark matter particle is presently ill-constrained. 
On cosmological scales, the density $\Omega_m$ of cold dark and baryonic matter combined ensures a three-flat Friedmann-Robertson-Walker (FRW) cosmology in the presence of a dark energy density $\Omega_\Lambda$, satisfying
\begin{eqnarray}
\Omega_\Lambda+\Omega_m=1, 
\label{EQN_3}
\end{eqnarray}
normalized to closure density $\rho_c=3H^2/8\pi G$, where $G$ is Newton's constant. Clustering of dark matter might be limited to the scale
of galaxy clusters, when its mass is extremely light on the order of \citep{van15b,van17a}
\begin{eqnarray}
m_{DM} \lesssim 10^{-30}\,\mbox{eV}.
\label{EQN_DM}
\end{eqnarray}
If so, the observed anamalous behavior in galaxy dynamics is different from conventional Newtonian gravitational dynamics
by baryonic matter content alone. A remarkable universality of (\ref{EQN_adS}) in galaxies across a broad range of surface brightness,
in spirals and elliptical galaxies alike \citep{lel17}, points to finite sensitivity to background cosmology. 

In this paper, we explain the prevalence of (\ref{EQN_adS}) in weak gravitational attraction in galaxy dynamics by perturbations of
inertia coevolving with cosmology, here confronted with recent data on galaxy rotation curves
covering a broad range of redshift from $z\sim 0$ \citep{lel16} to $z\sim 2$ \citep{gen17}. 

We recently proposed a holographic origin of inertia emergent from entanglement entropy derived from unitarity of their propagator \citep{van15a,van16,van17a,van17b}. Quite generally, describing particles by propagators starts with a choice of slicing -- a 3+1 foliation of spacetime in terms of Cauchy surfaces of constant coordinate time. 
In the presence of gravitation, these spacelike hypersurfaces of constant time may have apparent horizons, defined as the outermost marginally trapped surface describing a ``frozen light cone" that neither expands or contracts (turning points in surface area, \cite{bre88,yor89,wal91,coo92,coo00,tho07,van12}). As space-like intersections of null-surfaces, these horizons are gauge dependent. Equivalently, they are sensitive to the total gravitational field, as a covariant superposition of contributions of matter and acceleration of an observer \citep[e.g][]{fey03}. 
Examples are the {\em cosmological horizon} at Hubble radius
\begin{eqnarray} 
R_H=\frac{c}{H}
\label{EQN_RH}
\end{eqnarray}
in three-flat Friedmann-Robertson-Walker (FRW) cosmologies and apparent horizons surfaces in gravitational collapse signaling black hole formation \citep[e.g][]{bin15,far17a}. In the absence of matter, apparent horizon surfaces can arise from time-dependent boosts, i.e., in Rindler spacetimes, delineating a chart of Minkowski spacetime by a uniform gravitational field seen by a non-inertial observer at acceleration $\alpha$. The apparent horizon in Cauchy surfaces normal to the world-line of the accelerating observer is trailing at a constant distance 
\begin{eqnarray}
\xi = \frac{c^2}{\alpha}.
\label{EQN_xi}
\end{eqnarray}
In 3+1, these charts correspond to a local neighborhood outside the {\em event horizon} of a Schwarzschild black hole with similar thermodynamic properties \citep[e.g.][]{ati12}. These examples have in common apparent horizons in Cauchy surfaces, that carry entanglement entropy encoding particles positions at finite temperatures defined by surface gravity.

\begin{figure}[h]
\centerline{\includegraphics[scale=0.6]{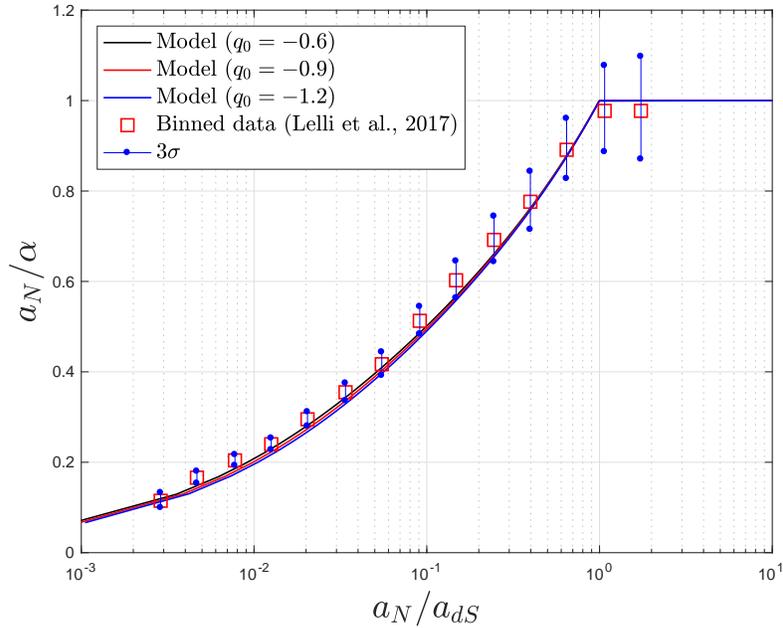}}
\vskip0.1in
\caption{The onset to weak gravity in high resolution galaxy rotation curves appears to be $C^0$ at Newtonian accelerations $a_N = a_{dS}$. A $C^0$ onset is expected for inertia $m$ of holographic origin, at collusion of apparent Rindler and cosmological horizons across which $m$ drops below the Newtonian value $m_0$. Plotted is $m/m_0=a_N/\alpha$ as a function of $a_N$ based on baryonic matter content normalized to the de Sitter acceleration $a_{dS}$. Binned data shown are accompanied by $3\sigma$ uncertainties, and continuous curves represent models for various deceleration parameters $q_0$ with Hubble parameter
$H_0 = 73$ km s$^{-1}$ Mpc$^{-1}$. (Adapted from \cite{van17a}).}
\label{figA}
\end{figure}
In FRW cosmologies, Cauchy surfaces of constant cosmic time are bounded by a cosmological horizon. Faced with both (\ref{EQN_xi}) and (\ref{EQN_RH}) in case of non-inertial observers, inertia from entanglement entropy of the {\em outermost} (closest to the observer) apparent horizon is perturbed when it colludes with the cosmological horizon at accelerations
\begin{eqnarray} 
\alpha=a_{dS}.
\label{EQN_C0}
\end{eqnarray}
The implied $C^0$ onset to weak gravity -- continuous with discontinuous derivatives -- sharply at (\ref{EQN_C0}) may be seen in recent high resolution data (Fig. \ref{figA}) of galaxies at redshifts $z\simeq0$ of \cite{lel16,lel17} ($H_0=73\,$km\,s$^{-1}$Mpc$^{-1}$, \cite{lel16a}), below which rotation velocities $V_c$ exceed their Keplerian values based on Newtonian acceleration $a_N$ to baryonic mass $M_b$ \citep{van17a}. For a mass $M_b = 10^{11}M_{11}$, the onset is at a transition radius 
\begin{eqnarray}
r_t = 4.7\,\mbox{kpc} \,M_{11}^{1/2} (H_0/H)^{1/2},
\label{EQN_rt}
\end{eqnarray} 
where $H_0=H(0)$. In $r < r_t$, gravity is conform Newton's theory. Following a $C^0$ onset, weak gravity in $r>r_t$ has radial accelerations $\alpha = V_c^2/r$
exceeding $a_N$ traced by light. The asymptotic regime $r>>r_t$ ($a_N << a_{dS}$) is commonly described by empirical relations of \cite{tul77} or, equivalently, \cite{mil83}.

We here consider sensitivity of weak gravity in galaxies to cosmology over an intermediate range of redshift at accelerations $a_N\lesssim a_{dS}$, 
recently covered by the sample of galaxy rotation curves of \cite{gen17} at the peak of galaxy formation, when $H(z)$ was about three times the present value $H_0$. 
The results provide an important test of weak gravity as shown in Fig. \ref{figA}.

\S2 describes three-flat Friedmann-Robertson-Walker (FRW) universes with dark energy and a cosmological distribution of cold dark and baryonic matter,
$\Omega_\Lambda$ and, respectively, $\Omega_M$. We identify dark energy $\Lambda$ with the square of a fundamental eigenfrequency of the cosmological horizon, and the resulting cosmological evolution is confronted with heterogeneous data on $H(z)$ covering redshifts $0<z<2$, including recent estimates 
of $H_0$ in surveys of the Local Universe. A holographic origin of inertia from entanglement entropy and its implications for weak gravity in galaxy dynamics below the de Sitter scale (\ref{EQN_adS}) are given in \S3. Based on Lorentz invariance of the cosmological horizon, previous results are generalized to circular orbits encountered in galaxy rotation curves.  This model is confronted with data of \cite{gen17} in \S4. In \S5, we summarize our findings.

\section{Cosmological evolution sans $H_0$ tension}

To facilitate an accurate and detailed confrontation with data on sensitivity of galaxy dynamics to background cosmology, we first 
model cosmologies with accelerated expansion. In the theory of general relativity, which assumes a classical
vacuum with no unseen small energy scales, dark energy was originally inferred from a deceleration parameter \citep{rie98,per99} 
\begin{eqnarray}
q= \frac{1}{2}\Omega_M - \Omega_\Lambda < 0,
\label{EQN_q}
\end{eqnarray}
where $q=-\ddot{a}a/\dot{a}^2 = -1 + (1+z) H^{-1}(z)H^\prime(z)$ with $a=a_0/(1+z)$ in the FRW line-element
\begin{eqnarray}
ds^2 = -dt^2 + a(t)^2\left( dx^2 + dy^2 + dz^2 \right)
\label{EQN_FRW}
\end{eqnarray} 
with $H(t)=\dot{a}(t)/a(t)$ in terms of the scale factor $a(t)$ measuring Hubble's law of galaxy velocities proportional to distance 
subject to (\ref{EQN_3}). 

The $H_0$ tension problem \citep{fre17} points to a dynamical dark energy \citep{van17b}, since the
relatively high value of $H_0$ observed in surveys of the Local Universe \citep{rie16} suggests fast evolution in $H(z)$ satisfying 
\begin{eqnarray}
H^\prime(0)\simeq0
\label{EQN_C2}
\end{eqnarray}
rather than $H^\prime(0)>0$ in $\Lambda$CDM.
  
Recently, we derived a dark energy $\Lambda=8\pi\rho_\Lambda$ \citep{van17a}
\begin{eqnarray}
\Lambda=\omega_0^2,~~\omega_0=\sqrt{1-q}H,
\label{EQN_AA}
\end{eqnarray}
in cosmological holography in terms of the fundamental eigenfrequency $\omega_0$
of the cosmological horizon, whereby
$\Lambda$ is {\em small, positive} and {\em dynamic}, pointed to by above mentioned observations. The finite eigenfrequency ($q<1$,
away from a radiation dominated epoch) is a direct consequence of compactness of the phase space of cosmological spacetime, set by the finite area of the
cosmological horizon. While early cosmology with (\ref{EQN_AA}) is very similar to $\Lambda$CDM, late time cosmology is 
different, marked by a relatively fast evolution of $H(z)$ and $q(z)$. While $q=1/2$ in the matter dominated era, (\ref{EQN_AA}) drives the 
deceleration to be {\em twice} the value (\ref{EQN_q}) in $\Lambda$CDM. In this cosmology, 
\begin{eqnarray} 
\Omega_\Lambda = \frac{1}{3}\left(1-q\right),~~\Omega_m=\frac{1}{3}\left(2+q\right)
\label{EQN_OM}
\end{eqnarray}
with an associated equation of state is $p_\Lambda=w\rho_\Lambda$, between pressure $p_\Lambda$ and energy density $\rho_\Lambda$
with $w=(2q-1)/(1-q)$, as opposed to $w\equiv -1$ in $\Lambda$CDM. 

Specifically, the FRW equations with (\ref{EQN_AA}) give a normalized Hubble parameter $h(z)=H/H_0$ with analytical solution
\begin{eqnarray}
h(z)  =  \sqrt{ 1 + \omega_m ( 6 z + 12 z^2 + 12 z^3 + 6 z^4 + \frac{6}{5}z^5 )}\,(1+z)^{-1},
\label{EQN_h}
\end{eqnarray} 
enabling a direct confrontation with data $(z,H(z))$ and $\Lambda$CDM, $h(z)=\sqrt{1-\omega_m + \omega_m (1+z)^3}$, over the two parameters $(H_0,\omega_m)$, where $\omega_m = \Omega_m(0)$ denotes the total of cold and baryonic cosmological matter density today. 

At high redshift, both (\ref{EQN_h}) and $\Lambda$CDM feature $h(z)\sim z^{3/2}$. 
At low $z$, their distinct behavior can be expressed in terms of $q(z)$ and $Q(z)=dq(z)/dz$, where $Q_0=Q(0)$
serves to indicate {\em fast} evolution characteristic for (\ref{EQN_C2}), satisfying $Q_0=(2+q_0)(1-2q_0)>2.5$ as opposed to 
{\em stiff} evolution in $\Lambda$CDM with $Q_0=(1+q_0)(1-2q_0)\lesssim1$. Distinct behavior at low $z$ between (\ref{EQN_h})
and $\Lambda$CDM can be seen explicitly from their leading order $z$-expansions
\begin{eqnarray}
h(z) = 1+(3\omega_m-1)z + O\left(z^2\right),~~h(z)=1+\frac{3}{2}\omega_mz +O\left(z^2\right),
\label{EQN_hz1}
\end{eqnarray}
giving, respectively, 
\begin{eqnarray}
H^\prime(0)\simeq0,~~H^\prime(0)\simeq 0.5\,H_0.
\label{EQN_hz2}
\end{eqnarray}
for a canonical value $\omega_m\simeq 0.3$.
A general Taylor series expansion in $(q_0,Q_0)$
\begin{eqnarray}
h(z) = 1 + (1+q_0)z+ \frac{1}{2}\left( Q_0 + q_0(1+q_0)\right)z^2 + b_3z^3 + \cdots
\label{EQN_T}
\end{eqnarray}
has, according to (\ref{EQN_h}) a radius of convergence $R=1$; for $\Lambda$CDM,
$R=(1-\omega_m)/\omega_m-1\ge1$ for $\omega_m\le1/3$. 
In applying (\ref{EQN_T}) for a model independent analysis of data $H(z)$ over some extended redshift range, 
$R=1$ implies that results on derivatives $h^{(n)}(z)$ ($n\ge1$) are valid for $0\le z<1$. 
This restriction is particularly relevant to graphs of $(q(z),Q(z))$ and their implications for estimates of $(q_0,Q_0)$ at $z=0$,
to be considered further below.

In considering (\ref{EQN_h}), our focus on sensitivity of estimates of $H_0$ to derivatives $H^\prime(0)$ in (\ref{EQN_hz2}) is different from consideration of regular perturbations of $\Lambda$CDM \citep[e.g][]{sie13,che17,far17} and its combinations with $\Lambda$CDM analysis of the CMB \citep{ade13,aub15} and Baryon Accoustic Oscillations \citep[e.g.][]{pod01}, by comparing fast versus stiff evolution according to (\ref{EQN_h}) and, respectively, $\Lambda$CDM.

Currently, $H(z)$ has been measured by various methods over an extended redshift range up to a redshift of about two \citep{sol17,far17}. 
Some of these data are not completely independent, which pertains to data from \cite{ala16} and \cite{pod01} at intermediate redshifts $0.38\le z\le 0.73$. 
In what follows, no correction for this interdependence has been made \citep{far17}. Because of the relatively minor number of data points involved (six), true uncertainties will be slightly greater than those quoted below. 

We analyse recently compiled heterogeneous data $\{z_k,H(z_k)\}$ over the extended redshift range $0<z_k<2$ that amply covers
the domain of convergence of the Taylor series expansion (\ref{EQN_T}). Estimates of cosmological parameters obtain by nonlinear model regression applied to (\ref{EQN_T}), (\ref{EQN_h}) and $\Lambda$CDM. The first serves as a model-independent analysis using truncated Taylor series to third and fourth order with no priors on any of the coefficients, to be compared with model fits over $H_0$ and one of the parameters $q_0,Q_0$ or $\omega_m$. In the $(q,Q)$-plane, graphs are plotted covering the data over $0\le z < 1$ conform $R=1$.

\begin{figure}[h]
\centerline{\includegraphics[scale=0.5]{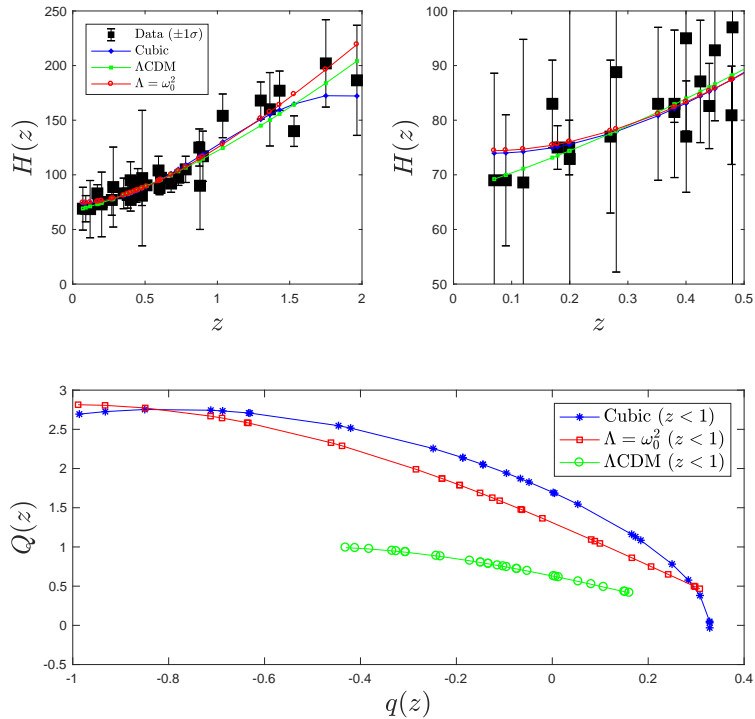}}
\vskip0.1in
\caption{(Top panels.) Fits by nonlinear model regression to a heterogeneous data set $(z_k,H(z_k))$ over $0<z<2$ applied
to the truncated Taylor series (\ref{EQN_T}), accelerated expansion by $\Lambda=\omega_0^2$ and $\Lambda$CDM. The latter
two show distinct behavior at $z$ close to zero with $H^\prime(0)\simeq0$, respectively, $H^\prime(0)>0$. (Bottom panel.)
In the $(q,Q)$-plane, results for the truncated Taylor series support $\Lambda=\omega_0^2$ but not $\Lambda$CDM.}
\label{figB}
\end{figure}

Fig. \ref{figB} shows results of our fits to data using the MatLab function {\em fitnml} with weights according to the inverse variance of data (cf. [5,18]) with otherwise default parameter settings. The results demonstrate the mechanism by which $\Lambda$CDM obtains relatively low values of $H_0$ by $H^\prime(z)>0$ everywhere, relative to $H_0$ in (\ref{EQN_h}) with $H^\prime(0)\simeq0$ in fast evolution driven by $\Lambda=\omega_0^2$. 

In the $(q,Q)$-plane, cubic fits are found to point to (\ref{EQN_h}) but not $\Lambda$CDM. 
The same is apparent in a tracking of of estimates of $\{H_0,q_0,Q_0\}$ over a running domain $D=[0,z]$ of data $(0<z<2)$ of
cubic fits with (\ref{EQN_h}) but not $\Lambda$CDM, shown in Fig. \ref{figB}.

\begin{figure}[h]
\centerline{\includegraphics[scale=0.5]{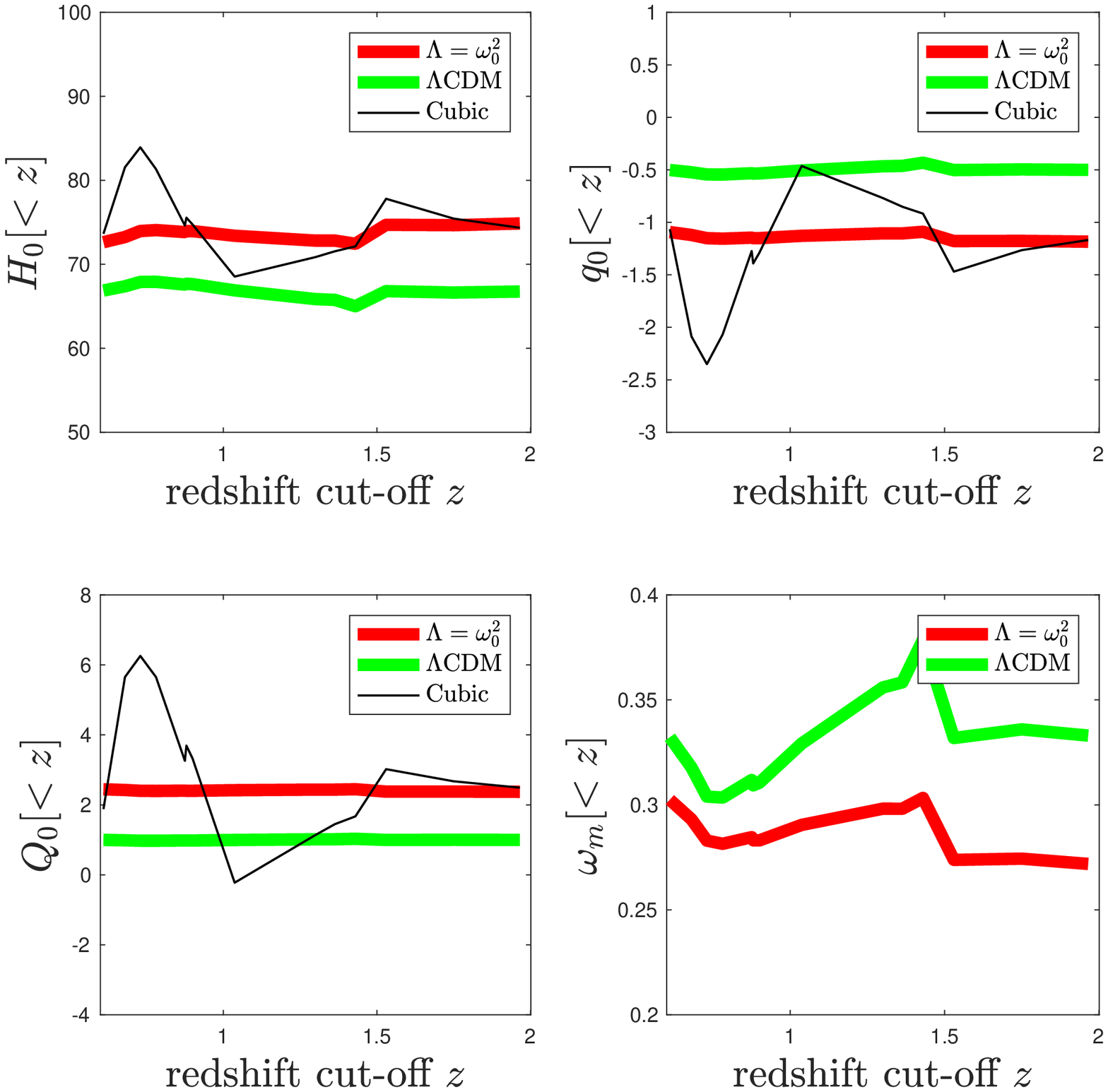}}
\vskip0.1in
\caption{Estimates $(H_0,q_0,Q_0,\omega_m)$ by nonlinear model regression as a function of domain
$[0,z]$. For $\{H_0,q_0,Q_0\}$, estimates by model-independent cubic fits (thin black) track model fits by $\Lambda=\omega_0^2$ (thick red) but not $\Lambda$CDM (thick green).}
\label{figC}
\end{figure}
    
Table 1 lists estimated values for $D=[0,2]$ with associated $1\,\sigma$ uncertainties. For the model fits with respect to
$(H_0,\omega_m)$, estimated uncertainties for $(q_0,Q_0)$ are derived according to the exact solutions $H(z)=H_0h(z)$
for (\ref{EQN_h}) and $\Lambda$CDM over $H_0$ and either one of $q_0$, $Q_0$ or $\omega_m$.

\begin{table}[h]
{\bf Table 1.} {Estimates of $(H_0,q_0,Q_0,\omega_m)$ with $1\,\sigma$ uncertainties by nonlinear model regression applied 
to the coefficients of the truncated Taylor series (\ref{EQN_T}) of cubic and quartic order, and to $(H_0,\omega_m)$ in
(\ref{EQN_h}) from $\Lambda=\omega_0^2$ and $\Lambda$CDM. $H_0$ is expressed in units of $\mbox{km~s}^{-1}\mbox{Mpc}^{-1}$.}
\center{\begin{tabular}{@{}lccccccc@{}}\mbox{}\\\hline\hline
	model & $H_0$ & $q_0$ & $Q_0$ & $\omega_m$ & $h^\prime(0)$ \\\hline
	Cubic                               & $74.4\pm4.9$  & $-1.17\pm0.34$  & $2.49\pm 0.55$ & - & -0.17 \\
	Quartic                             & $74.5\pm7.3$  & $-1.18\pm0.67$  & $2.54\pm 1.99$ &   & -0.18 \\
	$\Lambda=\omega_0^2$ & $74.9\pm2.6$  & $-1.18\pm0.084$ & $2.37\pm 0.073$ & $ 0.2719\pm0.028$ & -0.18 \\
	$\Lambda$CDM               & $66.8\pm1.9$  & $-0.50\pm0.060$ & $1.00\pm 0.030$ & $0.3330\pm0.040$ & 0.5 \\\hline
	\end{tabular}}\label{T1}\\
\end{table}
		
	Table 1 shows a three-fold consistency among model-independent cubic and quartic fits and the model fit to
	(\ref{EQN_h}). Estimates for $Q_0$ show the most stringent discrepancy with $\Lambda$CDM.
	Measured by the cubic fit, $\Lambda$CDM is inconsistent at $2.7\,\sigma$. 
	
	 The relatively large value $Q_0\simeq 2.5$ \citep{van15a} reported here is indicative of a near-extremal value of $H(z)$ today. This explains
	 our tension-free estimate of $H_0$ from cosmological data $\{z_k,H(z_k)\}$ with $H_0=73.24\pm1.74\,$km\,s$^{-1}$Mpc$^{-1}$ 
	 from surveys of the Local Universe [4], {\em higher} than $H_0$ in extrapolation to $z=0$
	 based on $\Lambda$CDM (with $H^\prime(z)>0$ everywhere). These considerations give quantitative support for the
	notion that dark energy may be dynamic rather than static. Combined with $H_0$ for $\Lambda=\omega_0^2$, we obtain
	\begin{eqnarray}
	H_0\simeq 73.75\pm 1.44\,\mbox{km\,s}^{-1}\mbox{Mpc}^{-1}.
	\label{EQN_H0a}
	\end{eqnarray}   

\section{Holographic inertia in weak gravity}

In unitary holography, inertia is of holographic origin emergent from entanglement entropy of an apparent horizon with thermodynamic potential \citep{van17a,van17b}
\begin{eqnarray}
U = mc^2.
\label{EQN_U}
\end{eqnarray}
In a Rindler spacetime at acceleration $\alpha$, it represents the integral of an entropic force integrated over separations $\xi=c^2/\alpha$ to a horizon, defined by $dU=-T_UdS=TdI_1$, where $T_U$ denotes the \cite{unr76} temperature {\em of the screen} and
\begin{eqnarray} 
I_1=2\pi \varphi
\label{EQN_I1}
\end{eqnarray}
the entanglement entropy encoding $\xi$ by total Compton phase.  (For a screen enveloping a black hole, (\ref{EQN_I1}) reduces to the Bekenstein-Hawking entropy of the event horizon \citep{bek73,haw75}.) Provided that $\xi < R_H$ $(\alpha > a_{dS})$, $U=m_0c^2$ of the Newtonian inertia $m_0$.

Essential to (\ref{EQN_U}) is the {\em non-local} and {\em instantaneous} nature of entanglement entropy associated with apparent horizons,
defined in Cauchy surfaces of given instances of eigentime, regardless of distances (\ref{EQN_xi}). 
As a response to acceleration, there is no known time delay between inertial reaction forces and acceleration. 

In a holographic representation of phase space, additional entanglement entropy resolves particles at finite angular resolution at opening angles $\Omega\simeq \pi \theta^2$ according to $\theta\simeq\lambda_C/r=2\pi/\Phi$, $\Phi=kr$, i.e., ${\Omega}/{4\pi} \simeq{\pi}/\Phi^{2}$, where $\lambda_C$ denotes the Compton wave length. Given $N=(1/4)Al_p^{-2}$ Planck sized surface elements for a screen of area $A=4\pi r^2$, each element projects out a small fraction of $4\pi$. The associated probability $p=\Omega/4\pi$ defines
\begin{eqnarray}
I_2 = - N\log p = 2\pi k_B r^2l_p^{-2} \log\Phi
\end{eqnarray} 
(up to an additional term independent of $M$), implications of which we shall not pursue further here.

The $C^0$ onset to weak gravity in Fig. \ref{figA} at (\ref{EQN_rt}) is a consequence of causality, as cosmological horizons impose a bound on charts of accelerating observers giving $U<m_0c^2$ whenever $\alpha < a_{dS}$. An effective mass ratio $\mu<1$, not unlike enhanced mobility, is apparent in accelerations that appear anomalously high,
\begin{eqnarray}
\alpha = \sqrt{\mu a_{dS}a_N} ~~(\alpha < a_{dS}).
\label{EQN_alpha}
\end{eqnarray} 
Here, $\mu$ is suitably expressed by a momentum average
\begin{eqnarray}
\mu = 2\left< B(p)\right>
\label{EQN_mu}
\end{eqnarray}
of the ratio $B(p)$ of two dispersion relations, of 
$\sqrt{k_B^2T_H^2 + c^2p^2}$ on the cosmological horizon at temperature
$T_H = {a}/{2\pi}$ at internal surface gravity $a=\frac{1}{2}(1-q)a_{dS}$ and $\sqrt{\hbar^2\Lambda + c^2p^2}$ in 3+1 spacetime within, i.e.,
$B(p) = {\sqrt{k_B^2T_H^2+c^2p^2}}/{\sqrt{\hbar^2\Lambda+c^2p^2}}$ with Boltzmann constant $k_B$ and reduced Planck constant $\hbar$.

A thermal average by a Boltzmann factor $e^{-E/T}$, $E=\sqrt{\hbar^2\Lambda+c^2p^2}-\hbar\sqrt{\Lambda}$ gives 
\begin{eqnarray}
\left<B(p)\right>_y = \frac{1}{W}\int_0^\infty\sqrt{\frac{1+x^2}{A+x^2}}e^{-e/y}x^2dx,
\label{EQN_Bp}
\end{eqnarray}
$A={16\pi^2}/(1-q)$, parameterized by 
\begin{eqnarray}
y = y_0 \left(\frac{1-q}{2}\right)^{-\gamma},~~y_0 = \frac{a_N}{a_{dS}}.
\label{EQN_y0}
\end{eqnarray}
The average (\ref{EQN_Bp}) is essentially continuous at $y_0=1$, i.e.: $2\left< B(p)\right>_y\simeq1$, $y=((1-q)/2)^{-\gamma}$, where the emperical value $\gamma=0.535$ serves to cover the range $-1\lesssim q \lesssim 0.5$.

In the classical de Sitter limit described by $\Omega_\Lambda=1$ and $\Omega_M=0$, the cosmological horizon is a Lorentz invariant: inertial observers, regardless of boost and location, all agree on the same Hubble radius $R_H$ to their cosmological horizons. Since inertia is defined by local ({\em instantaneous}) curvature of particle trajectories, (\ref{EQN_C0}) trivially follows as the condition for a $C^0$ perturbation according to our previous analysis, after transforming away any instantaneous velocity by a boost. In a non-de Sitter cosmology such as ours, the same still applies in case of circular orbits, as tangential boosts leave invariant spacelike distances in orthogonal, radial directions associated with the orbital radius of curvature. An observer boosted along sees a radial
acceleration, like that imparted by a rope or Newtonian gravitational force on a given cosmological background (\ref{EQN_FRW}). 
Essential for inertia associated with local curvature of particle trajectories to be emergent is, as mentioned above, its holographic origin 
in apparent horizons and the non-local and instantaneous nature of entanglement entropy therein.

\begin{figure}[h]
\centerline{\includegraphics[scale=0.6]{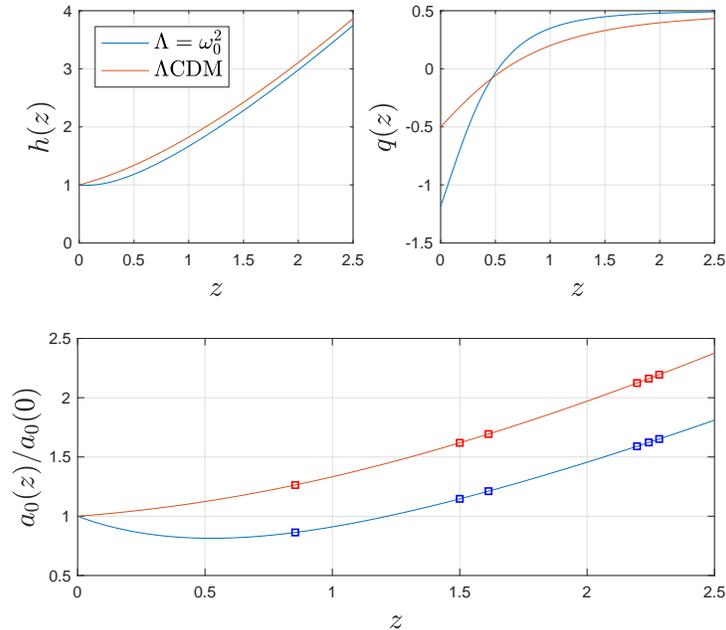}}
\caption{(Top panels.) Following the results of Table 1, shown are $h(z)$, $q(z)$ and Milgrom's parameter $a_0(z)$
for cosmological evolution by $\Lambda=\omega_0^2$ and $\Lambda$CDM. $a_0(z)$ varies only moderately, far 
less than the change in $H(z)$ by a factor of about three over the redshift range $0\le z \le 2.5$. Squares refer
to the location of the galaxies listed in Table 1.}
\label{figa0}
\end{figure}

In galaxy rotation curves of \cite{lel16}, (\ref{EQN_alpha}) is found to be in agreement covering the onset to weak gravity sharply at (\ref{EQN_C0}) down to $\alpha << a_{dS}$ (Fig. \ref{figA}). The latter is described by the empirical baryonic \cite{tul77} relation, equivalent to \cite{mil83} law with
\begin{eqnarray}
a_0 = \frac{\omega_0}{2\pi} ~~(\alpha << a_{dS}).
\label{EQN_a0}
\end{eqnarray}
Fig. \ref{figa0} shows $a_0(z)$ on the cosmological backgrounds of Table 1. While $H(z)$ varies by a factor of about three over $0\le z \le 2$,
$a_0(z)$ varies merely by approximately 50\% in case of $\Lambda=\omega_0^2$. Consequently, and especially so for the intermediate range of weak gravity
when $a_N/a_{dS}\lesssim1$, redshift sensitivity of inertia derives in (\ref{EQN_y0}) primarily through normalization by $a_{dS}$.

\section{Weak gravity in rotation curves at $z\sim2$}

Recently, \cite{gen17} reported rotation curves of a sample of six galaxies at intermediate redshifts with rotation velocities $V_c$  reported at a specific radius $R_{1/2}$, as a fit to the H-band half-light radius $R_h$ in a cosmology with $\omega_m=0.3$
and Hubble parameter $H_0=70$ km s$^{-1}$ Mpc$^{-1}$ (Table 1).

For a detailed confrontation of (\ref{EQN_alpha}) with data on spiral galaxies with a total baryonic mass $M_b$, consisting of a baryonic 
mass $M_{bulge}$ in the central bulge and a mass $M_b-M_{bulge}$ in stars and gas in an extended disk,  we shall use the following 
equivalent point mass 
\begin{eqnarray}
M_b^\prime = f_a\left(M_{bulge}+f_b(M_b-M_{bulge})\right)
\label{EQN_Mb}
\end{eqnarray}
where $f_b = 1/\chi$ converts (an infinitely thin) disk mass to equivalent point mass \citep{mcg12} and $f_a$ is the fraction of baryonic mass at $R_{1/2}$ to total mass. In \cite{gen17}, Extended Data Fig. 4 on GS4 43501 shows $f_a\simeq 0.6$.

By continuity in the onset to weak gravity at $\alpha=a_{dS}$ (\ref{EQN_rt}), we have 
\begin{eqnarray}
y_{0,1/2}=\left(\frac{a_N}{a_{dS}}\right)_{1/2} =\left( \frac{r_t}{R_{1/2}}\right)^2
\end{eqnarray}
whereby (\ref{EQN_alpha}) takes the form
\begin{eqnarray}
\frac{\alpha}{a_{dS}} = \sqrt{ \mu} \frac{r_t}{R_{1/2}}
\label{EQN_alpha2}
\end{eqnarray}
In weak gravity $\alpha \le a_{dS}$ away from the asymptotic limit $\alpha<<a_{dS}$, anomalous behavior in galactic dynamics may be expressed by an apparent dark matter fraction
\begin{eqnarray}
f_{DM}^\prime = \frac{\alpha - a_N}{\alpha},
\label{EQN_fDM}
\end{eqnarray}
conform $f_{DM}$ in \cite{gen17}. In Table 1, $f_{DM}$ of \cite{gen17} based on $H_0=70$ km$^{-1}$ s$^{-1}$ Mpc$^{-1}$ are not
adjusted to our value of $H_0\simeq 73$ km$^{-1}$ s$^{-1}$ Mpc$^{-1}$, that would produce slightly higher values by about 4\%.

\begin{figure}[h]
\begin{center}\includegraphics[scale=0.5]{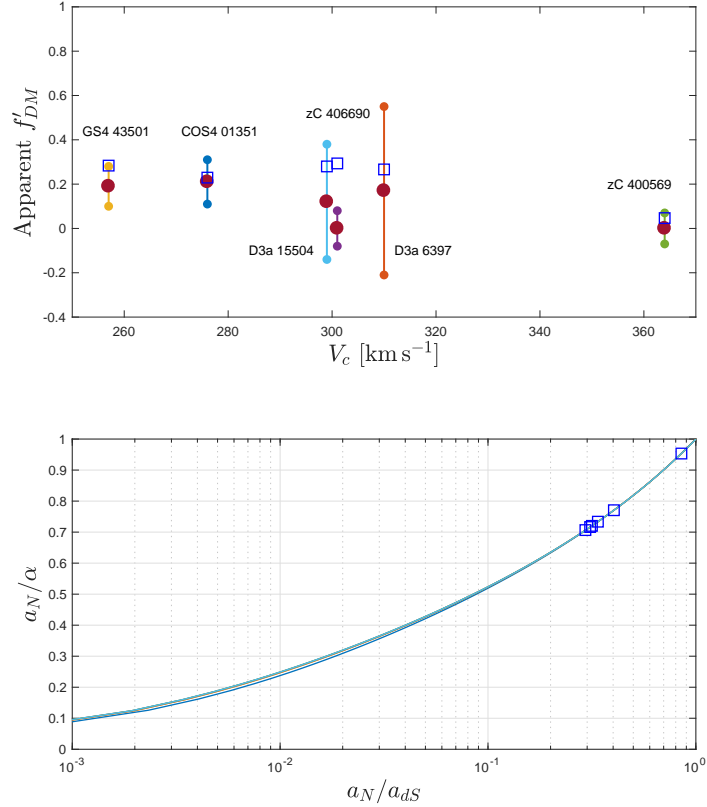}\end{center}\vskip0.1in
\caption{(Upper panel.) Apparent dark matter fractions $f_{DM}^\prime$ (blue squares) based on (\ref{EQN_alpha}) agree with observational data on $f_{DM}$ except for cZ 406690 ($V_c=301$ km/s). (Lower panel.) Squares show the clustering of $a_N/a_{dS}$ in the Genzel et al. sample (Table 2) close to the $C^0$ onset to weak gravity at $a_N/a_{dS}=1$ (Fig. \ref{figA}). Continuous lines are model curves, one for each galaxy redshift.}
\label{FRW}
\end{figure}

Fig. \ref{FRW} shows the results on (\ref{EQN_fDM}). Based on $r_t/R_{1/2}$, the \cite{gen17} probes weak gravity, not the asymptotic regime (\ref{EQN_a0}), recently emphasised by \cite{mil17} and shown here in the lower panel of Fig. \ref{FRW}. 

\begin{table}[h]
{\bf Table 2.} {Weak gravity analysis on apparent $(f_{DM}^\prime)$ versus observed $(f_{DM})$ dark matter fractions in rotation curves of the \cite{gen17} galaxies with equivalent central baryonic mass $M_b^\prime$ according to (\ref{EQN_Mb}) and rotation velocities $V_c$ [km\,s$^{-1}$] at radius $R_{1/2}$ [kpc].}
\center{{\begin{tabular}{@{}lcccccccccc|cl@{}}
\mbox{}\\\hline\hline
	Galaxy &  $M_b^\prime/M_b$ & $R_{1/2}$ & $z$ & $h(z)$ & $q(z)$ & ${a_0(z)}/{a_0(0)}$ & $M_{b,11}$ & $r_t/R_{1/2}$ & $\mu$ & $f^\prime_{DM}$ & $V_c$ & $f_{DM}$ (95\% cl)\\
	\hline
	COS4 01351 & 0.86 & 7.3 & 0.854  & 1.5986 & 0.0853 & 0.8632 & 1.7  & 0.6351 & 0.6793 & 0.2294 &276 & 0.21$\pm0.1$\\
	D3a 6397      & 0.83 & 7.4 & 1.500  & 2.2883 & 0.2957 & 1.1468 & 2.3  & 0.5812 & 0.6276 & 0.2663 & 310 & 0.17 $(<0.38)$\\
	GS4 43501   & 0.82 & 4.9 & 1.613  & 2.4246 & 0.3170 & 1.2116 &  1.0  & 0.5568 & 0.6040 & 0.2835 & 257 & 0.19 $(\pm0.09)$\\
	zC 406690    & 0.78 & 5.5 & 2.196  & 3.1903 & 0.3996 & 1.5902 & 1.7  & 0.5424 &0.5893 & 0.2934 & 301 & 0 $(<0.08)$\\
	zC 400569    & 0.83 & 3.3 & 2.242 & 3.2531 & 0.3919 & 1.6227 &  1.7  & 0.9231 & 0.9377 & 0.0467 & 364 & 0 $(<0.07)$\\ 
	D3a 15504    & 0.87 & 6 & 2.383 & 3.4537 & 0.4184 &  1.6519 & 2.1 & 0.5623 & 0.6997 & 0.2798 & 299 & 0.12 $(<0.26)$\\ 
\end{tabular}}}
\label{Table_val}
\end{table}

Fig. \ref{FRW} shows general agreement of $f_{DM}^\prime$ and $f_{DM}$, except for cZ 4006690 ($V_c=301$ km s$^{-1}$). However, its observed rotation curve is strongly asymmetric, that may entail systematic errors different from the remaining galaxies with essentially symmetric rotation curves. Further in light of the agreement of theory with rotation curves of \cite{lel16} at zero redshift \citep{van17a}, there appears to be no tension in apparent $f_{DM}^\prime$ and observed $f_{DM}$ in rotation curves over a broad range of redshifts, confirming (\ref{EQN_rt}) and the overall sensitivity of galactic dynamics to background cosmology, mostly so to $H(z)$ through normalisation of $a_N$ by $a_{dS}$.

\section{Conclusions}

We report on a probe of weak gravity probed by galaxy dynamics in an evolving background cosmology, of
anomalous behaviour over a broad range of accelerations and redshifts, i.e.:
$0.003\lesssim a_N/a_{dS}\lesssim1$ (Fig. \ref{figA}) and $0\lesssim z\lesssim2.5$ (Fig. \ref{FRW}).
We identify an onset at $a_N=a_{dS}$ that appears to be $C^0$,
pointed to by high resolution data of galaxies about redshift $z\sim0$ (Fig. \ref{figA}).
 
We explain anomalous behaviour with a holographic origin of inertia emergent from entanglement entropy,
perturbed at small accelerations in the face of a finite cosmological horizon.
Based on Lorentz invariance of the cosmological horizon, we apply our theory of inertia in radial motions in Minkowski spacetime
to circular motions in galaxy rotation curves in a three-flat cosmological background.

A $C^0$ onset to weak gravity sharp at $a_N=a_{dS}$ points to a sensitivity of weak gravity to cosmological evolution. Weak gravity
is primarily a function primarily of $a_N/a_{dS}$ with a moderate redshift dependence in asymptotic behaviour parameterised by
$a_0(z)$, limited to 65\% for the intermediate redshift range considered here, even as $H(z)$ varies by a factor of about three. 

For a cosmology with $\Lambda=\omega_0^2$ favoured by Table 1, $a_0(0)\simeq 1.66\times 10^{-8}$cm\,s$^{-2}$ 
in (\ref{EQN_a0}) is consistent with data (Fig. \ref{figA}), but appears to be somewhat higher than early estimates of
$a_0(0)\simeq 1.2\times 10^{-8}$cm\,s$^{-2}$ as averages of samples with considerable scatter \citep{beg91}
or best-fit estimates using interpolation functions \citep[e.g.][]{fam12} with systematic uncertainties of $\sim$20\% \citep{lel16}. 
Perhaps this also reflects our modeling of weak gravity by $2<B(p)>_y$, whose $C^0$ onset is ignored in canonical smooth interpolations of data
and whose converge to $\alpha=\sqrt{a_0a_N}$ by $\alpha=\sqrt{a_0a_N}(1+O(x))$ ($x=a_N/a_0<<1$) is faster than by $\alpha=\sqrt{a_0a_N}(1+O(\sqrt{x}))$ obtained with canonical interpolation functions. 
A detailed study of this discrepancy will be pursued elsewhere. 

While current rotating curve data show remarkable improvements in resolution, they remain
inadequate to meaningfully constrain $q(z)$ directly by $a_0(z)$ through (\ref{EQN_a0}) at any particular redshift $z$ at
accelerations $\alpha << a_{dS}$. Even so, the broad redshift range covered by the galaxies at
$z\sim0$ (Fig. \ref{figA}) and $z\sim2$ (Fig. \ref{FRW}) strongly support a moderate variation of $a_0$ 
\citep{mil17}, here identified with a relatively flat behaviour of $\omega_0(z)= \sqrt{1-q(z)}H(z)$ in (\ref{EQN_a0}).

A $C^0$ onset to weak gravity is a specific prediction of holographic inertia. This type of onset
is at odds with any smooth distribution of dark matter clustering on galactic scales. Instead, we propose the putative
dark matter particle to be extremely light (\ref{EQN_DM}), allowing clustering on the scale of galaxy clusters but not individual galaxies.

{\bf Acknowledgements.} The author gratefully acknowledges constructive comments from R.H. Sanders, which greatly improved presentation of this manuscript. This research is supported in part by the National Research Foundation of Korea (No. 2015R1D1A1A01059793 and 2016R1A5A1013277).

\end{document}